# Computations Meet Experiments to Advance the Enzymatic Depolymerization of Plastics One Atom at a Time

Francesco Colizzi[1,2],* Paula Blázquez-Sánchez[3],* Giovanni Bussi[4], Isabelle André[5], Federico Ballabio[6], Thomas Bayer[35], Federica Bertocchini[7], Erik Butenschön[8], Emanuele Carosati[9], Alessia De Piero[9], Ania Di Pede-Mattatelli[2,10], Leonardo Faggian[11], Lorenzo Favaro[11,12,13], Pedro Alexandrino Fernandes[14], Alfonso Ferretti[15,16], Peter Fojan[17], Andreas Gagsteiger[18], Eva García-Ruiz[19], Lucia Gardossi[9], Toni Giorgino[6], Norbert Graefe[3], Jonas Gunkel[3], Diego J. Jiménez[21], Malene Billeskov Keller[22], Georg Künze[23,24,25], Sierin Lim[26], Guy Lippens[5], Miguel A. Maria-Solano[2], Karolina Mitusińska[20], Vicent Moliner[27], Gianluca Molla[28], Marthinus Wessel Myburgh[11,12,13], Nicolas Panel[5], César A. Ramírez-Sarmiento[29,30], Dominique Rocher[11,12,13], Birgit Strodel[31,32], Belén Taroncher Ruiz[33], Onur Turak[18], Cristiano Varrone[34], Daniele Vezzini[11,12], Peter Westh[22], Wolfgang Zimmermann[8]

(1) Molecular Ocean Lab, Institute for Advanced Chemistry of Catalonia (IQAC–CSIC), Carrer de Jordi Girona 18-26, 08034 Barcelona, Spain

(2) Molecular Ocean Lab, Institute of Marine Sciences (ICM–CSIC), Passeig Marítim de la Barceloneta 37-49, 08003 Barcelona, Spain

(3) Institute of Drug Discovery, Leipzig University, 04103 Leipzig, Germany

(4) Scuola Internazionale Superiore di Studi Avanzati (SISSA), via Bonomea 265, 34136, Trieste, Italy

(5) Toulouse Biotechnology Institute, TBI, Université de Toulouse, CNRS, INRAE, INSA, Toulouse, France. 135, avenue de Rangueil, F-31077 Toulouse Cedex 04, France.

(6) Istituto di Biofisica (IBF-CNR), Consiglio Nazionale delle Ricerche, via Celoria 26, 20133, Milano, Italy

(7) Plasticentropy France, 28 Rue de Thiers, 51100 Reims, France

(8) Institute of Analytical Chemistry, Leipzig University, 04103 Leipzig, Germany

(9) Dipartimento di Scienze Chimiche e Farmaceutiche, Università degli Studi di Trieste, Via Licio Giorigieri 1, 34127, Trieste, Italy

(10) Graduate Programme in Biotechnology, Universitat de Barcelona, Barcelona, Spain

(11) Waste to Bioproducts Lab – DAFNAE, University of Padova, Viale dell'Università 16, 35020, Legnaro (PD), Italy

(12) Microbiology Department, University of Stellenbosch, Stellenbosch, South Africa

(13) Urobo Biotech,15 De Beer Street,Stellenbosch, South Africa

(14) LAQV@Requimte, Faculty of Sciences, University of Porto, Rua do Campo Alegre S/N, 4169-007 Porto, Portugal

(15) Scuola Normale Superiore, Piazza Dei Cavalieri 7, I-56126 Pisa, Italy

(16) Istituto Nazionale di Fisica Nucleare (INFN), Largo Pontecorvo 3, I-56127 Pisa, Italy

(17) Department of Materials and Production, Aalborg University, Fibigerstræde 16, 9220 Aalborg, Denmark

(18) Department of Biochemistry, University of Bayreuth, Universitaetsstrasse 30, 95447 Bayreuth, Germany

(19) Instituto de Catálisis y Petroleoquímica, ICP-CSIC, C/Marie Curie, 2, 28049, Madrid, Spain

(20) National Research Council of Italy (CNR)-IOM c/o Scuola Internazionale Superiore di Studi Avanzati (SISSA), via Bonomea 265, Trieste 34136, Italy

(21) Biological and Environmental Sciences and Engineering Division (BESE), King Abdullah University of Science and Technology (KAUST), Thuwal, 23955-6900, Kingdom of Saudi Arabia

(22) Department of Biotechnology and Biomedicine, Technical University of Denmark, DK-2800 Lyngby, Denmark

(23) Institute of Drug Discovery, Leipzig University, 04103 Leipzig, Germany

(24) Center for Scalable Data Analytics and Artificial Intelligence, Leipzig University, 04105 Leipzig, Germany

(25) Interdisciplinary Center for Bioinformatics, Leipzig University, 04107 Leipzig, Germany

(26) School of Chemistry, Chemical Engineering and Biotechnology & Nanyang Environment and Water Research Institute (NEWRI), Nanyang Technological University, Singapore 637457

(27) Institute of Advanced Materials. Universitat Jaume I. 12006 Castellón. Spain.

(28) Dipartimento di Biotecnologie e Scienze della Vita, Università degli Studi dell'Insubria, via J.H. Dunant 3, 21100 Varese, Italy

(29) Institute for Biological and Medical Engineering, Pontificia Universidad Católica de Chile, Santiago, Chile

(30) ANID, Millennium Institute for Integrative Biology (iBio), Santiago, Chile

(31) Institute of Biological Information Processing (IBI-7: Structural Biochemistry), Forschungszentrum Jülich, Jülich, Germany

(32) Institute of Theoretical and Computational Chemistry, Heinrich Heine University Düsseldorf, Düsseldorf, Germany

(33) Plastic Research Institute (AIMPLAS), Carrer de Gustave Eiffel 4, 46980 Valencia, Spain

(34) Department of Chemistry and Bioscience, Section of Bioresources and Process Engineering, Aalborg University, Fredrik Bajers Vej 7H, 9220 Aalborg, Denmark

(35) Department of Biotechnology & Enzyme Catalysis Institute of Biochemistry, University of Greifswald, Felix-Hausdorff-Str. 4, 17487 Greifswald, Germany

**Corresponding authors:** colizzi@csic.es; paula.blazquez@uni-leipzig.de

**Abstract**

Plastics are essential to modern life, yet poor disposal practices contribute to low recycling rates and environmental accumulation—biological degradation and by-product reuse offer a path to mitigate this global threat. This report highlights key insights, future challenges, and research priorities identified during the CECAM workshop *"Computations Meet Experiments to Advance the Enzymatic Depolymerization of Plastics One Atom at a Time,"* held in Trieste from May 6–8, 2025. The workshop brought together an interdisciplinary community of scientists focused on advancing the sustainable use of plastics through enzyme-based degradation. A key point from the discussions is that many bottlenecks in enzymatic recycling arise not only from process engineering challenges, but also from a limited understanding of the underlying molecular mechanisms. We argue that constraints on economic viability and sustainability (e.g., harsh solvents, high temperatures, substrate crystallinity, pretreatments) can—and should—be addressed directly through enzyme design, provided these factors are understood at the molecular level, in synergy with process optimization. For this, it is essential to rely on the integration of experimental and computational approaches to uncover the molecular and mechanistic basis of enzymatic plastic degradation. We highlight how the small-format structure of the workshop, in line with the usual CECAM format, fostered a collaborative, friendly, and relaxed atmosphere. We hope this report encourages future initiatives and the formation of shared consortia to support an open, collaborative, and bio-based plastic recycling community.

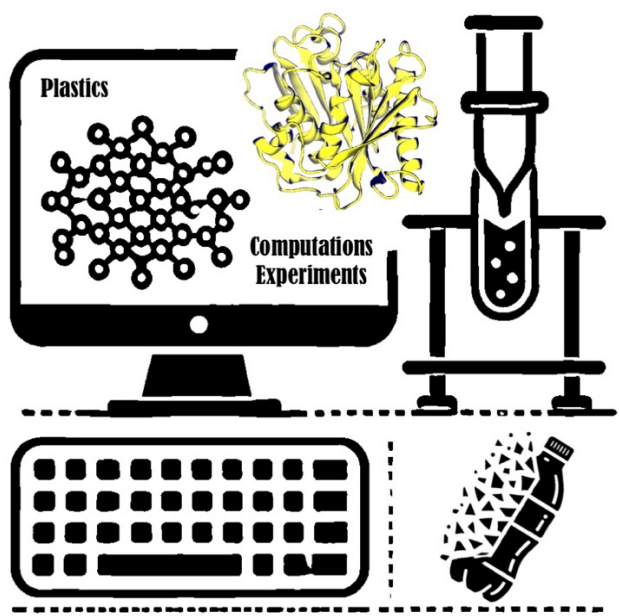

**Introduction**

Plastics are indispensable in modern life—with annual production exceeding 500 million metric tons and projected to reach 700 million by 2050.[1] Yet, the improper disposal of plastic waste, especially single-use materials such as polyethylene terephthalate (PET), has contributed to global recycling rates remaining below 30%. As a result, the majority of plastic waste is currently incinerated, landfilled, or leaks into ecosystems at staggering rate of 50 million metric tons each year.[2]

Sustainable biological approaches, particularly enzyme-based plastic degradation, are emerging as promising complements to conventional mechanical and chemical recycling of several plastics, including PET, polyethylene (PE), polyurethane (PU), polylactic acid (PLA), polybutylene succinate (PBS), among others, with current efforts worldwide overwhelmingly focused on enzymatic PET degradation.[3–5] These biocatalytic strategies offer a pathway toward a circular plastic economy.[6,7] However, despite growing industrial interest, enzyme-based depolymerization remains largely limited to laboratory or pilot-scale settings. Only a small fraction of existing plastic waste is amenable to enzymatic breakdown, and the development of scalable, cost-effective processes is still in progress.[3]

While technical bottlenecks are often attributed to process engineering, and innovative solutions are emerging,[8] this report argues that critical limitations also stem from an incomplete understanding of the fundamental molecular mechanisms governing enzymatic depolymerization. Non-enzymatic parameters—such as pH, temperature, buffer composition and concentration, substrate crystallinity and loading, as well as agitation and mixing— profoundly affect enzyme performance and influence structure–function relationships. A deeper understanding of the molecular features linking these parameters to enzyme activity is essential to overcome current barriers and accelerate the practical deployment of plastic-degrading enzymes, in synergy with process innovations.

This report summarizes key insights and future challenges identified during the CECAM workshop *"Computations Meet Experiments to Advance the Enzymatic Depolymerization of Plastics One Atom at a Time,"* held in Trieste from May 6–8, 2025. The workshop convened a diverse, interdisciplinary community of academic and industrial researchers working at the interface of molecular biophysics, computational chemistry, structural biology, enzymology, molecular design, polymer chemistry, and bioprocess engineering. A central theme was the integration of computations and experiments to reveal the molecular and mechanistic basis of enzymatic plastic degradation.

Key topics included enzyme structure–function relationships, protein design and engineering, discovery of novel biocatalysts, process upscaling, development of sustainable polymers, and the coupling of enzyme engineering with process optimization. Discussions were steered toward a molecular focus, reaffirming the potential of microbial enzymes—such as PET hydrolases, amidases, and hexamerins from wax worms—for degrading plastic polymers under mild, environmentally compatible conditions.

*This report aims to highlight challenges and perspectives raised during the event recognizing that numerous other contributions outside this discussion are not covered here.*

**The Long Way from Enzymatic Degradation to Scalable Industrial Application**

A major focus of the workshop was the challenge of transitioning enzymatic degradation from research settings to scalable industrial applications. Presentations addressed the engineering of microbial PET hydrolases into high-performance biocatalysts through computational design, protein engineering, and machine learning.[4] These redesigned enzymes show increased thermal stability and activity, enabling efficient depolymerization of post-consumer plastic waste (W. Zimmermann). Enzymatic recycling offers an attractive alternative to mechanical and chemical approaches, with real-world case studies highlighting both the promise and the limitations of integrating these technologies into existing waste management systems (B. Taroncher Ruiz).

Beyond monomer recovery, engineered microbial platforms are being developed to upcycle plastic waste into new, more recyclable bio-based polymers.[9] A key step toward this goal is the design of one-pot processes, in which enzyme production and PET degradation occur in a single step—eliminating the need for costly and time-consuming purification procedures.[10] This would lay the basis for the development of consolidated bioprocesses where plastic waste can be depolymerized and potentially bio-upcycled within the same reactor. At pilot scale, such integrated one-pot processes have shown promise but also revealed new bottlenecks, particularly in pretreatment steps, enzyme production costs (due to low protein yields and the use of costly antibiotics and gene expression inducers), and NaOH consumption. These challenges are currently being addressed using AI-assisted tools for process optimization (C. Varrone).

Novel computational frameworks are advancing the rational design of plastic-degrading enzymes. Hybrid strategies that combine electrostatics-guided modeling with quantum mechanics/molecular mechanics (QM/MM) simulations are now being used to predict rate-limiting steps, effect of pH, and identify key catalytic residues across a range of substrates, including PET, PU, and nylon (V. Moliner; P.A. Fernandes).[11,12] Structural and functional optimization of PET hydrolases have been advanced through Rosetta-based redesign and molecular dynamics (MD) simulations, as shown by the enhanced thermostability and up to 120-fold increase in catalytic activity of engineered variants of the Polyester Hydrolase Leipzig 7 (PHL7)[13] (G. Künze). Phylogenetic analysis and Rosetta-based design tools (e.g., FuncLib)[14] have been used to engineer PHL7 for improved degradation performance on PET, PBS, and PLA. Experimental validation via electrochemical impedance spectroscopy confirmed enhanced enzymatic activity and stability (J. Gunkel). In parallel, targeted studies explored engineered loop variants to assess their impact on PET hydrolysis efficiency (N. Graefe).

Deep-learning-guided sequence redesign is being used to generate PET hydrolase variants with up to 120-fold higher expression yields in *E. coli* and increased thermal stability compared to wild-type enzymes. Although only a small fraction (<5%) of these variants exhibit even weak PETase activity, the approach highlights both the promise of data-driven enzyme engineering and the inherent complexity of enhancing PET hydrolase activity (C. Ramírez-Sarmiento). Minimalist, *de novo*–designed protein scaffolds with implanted catalytic sites are being explored to overcome the structural limitations of naturally evolved folds.[15] These efforts aim to enable the rational design of new enzyme classes specifically tailored to degrade persistent plastics such as crystalline PET. This approach also allows efficient escape from the hydrolase architecture and explores new folds, other than the classical alpha/beta hydrolase fold,[5] that might be better suited for enzymatic activity against non-natural substrates (P. Fojan). Machine learning is also accelerating enzyme discovery and optimization, particularly through the development of tools for predicting thermostability and prioritizing promising variants (A. Di Pede-Mattatelli).

Enhanced-sampling MD simulations are unlocking detailed quantification of the energetic landscape associated with accessing productive catalytic states on both amorphous and crystalline PET substrates, including the energy cost of chain extraction from a minimal, idealized polymer matrix (A. Di Pede-Mattatelli). MD simulations and protein network theories are providing new insights into catalytic mechanisms and allosteric communication in both natural and engineered hydrolases, identifying mutation hotspots for further optimization (M.A. Maria-Solano). At the atomic level, simulations of polyester hydrolysis are informing detailed mechanistic models for PETase and MHETase, as well as expanding to PU-active hydrolases (P.A. Fernandes). Multiple MD-based studies are investigating the PETase binding process—from initial surface adsorption to substrate threading and active-site engagement—highlighting how binding preferences and catalytic efficiency differ between amorphous and crystalline PET (P.A. Fernandes; B. Strodel).[16,17] These studies also inform targeted mutations aimed at enhancing PET adsorption (B. Strodel), and they analyze how various PETases and carbohydrate-binding modules interact with different PET morphologies to better understand substrate preferences. Subsequent QM/MM simulations elucidate the effects of mutations near the active site on the enzymatic activity of different Leaf and branch Compost Cutinase (LCC) and PHL7 variants (B. Strodel). MD simulations are guiding the design of thermostable PETase variants by revealing temperature-dependent dynamics (N. Jeyaraj Suja). Additionally, new tools analyzing solvent and small molecule transportation through enzyme tunnels and cavities provide insights into internal architecture for targeted protein engineering (K. Mitusinska). Complementing these efforts, Machine learning potentials based on optimized DFT-D calculations are improving the accuracy of MD simulations of water, allowing better modeling of enzyme environments (A. Ferretti).

Nuclear magnetic resonance (NMR) spectroscopy is increasingly used to investigate enzyme–polymer interactions and to monitor product formation in real time. Insights from both solution and solid-state NMR offer multiscale understanding of catalytic interfaces, capturing dynamic binding modes and catalytic conformations. Solution NMR studies track mutation-induced changes in the PETase catalytic triad and correlate them with enhanced enzymatic activity (G. Lippens).[18] By locking the enzyme-polymer system in a trehalose glassy matrix, solid-state NMR measurements reveal the atomic-level mechanisms underlying enzymatic PET degradation throughout the reaction.[19] These studies, combined with MD simulations, allow monitoring and quantifying specific enzyme–PET interactions during degradation (E. Butenschön).

**Expanding the Enzyme Toolbox Beyond PET hydrolases**

A further theme of the workshop was the discovery and development of enzymatic activities beyond PET. Oxidative enzymes derived from wax worms have shown activity on polyethylene, opening a new avenue for polyolefin degradation (F. Bertocchini).[20] Recombinant enzymes from wax worm saliva have been investigated for their depolymerization potential, with experimental assays and preliminary molecular docking studies shedding light on their interactions with plastic substrates (M. Boukhallat). Novel amidases and cutinases capable of acting on polyester–polyurethane and nylon-6 have also been identified and engineered for enhanced performance (T. Bayer). Evolutionary strategies, including ancestral sequence reconstruction combined with directed evolution, have produced PET hydrolases with improved activity, expression, and stability tailored for integration into consolidated bioprocesses (E. Garcia Ruiz). Exploration of marine and saline environments has led to the identification of salt-tolerant bacterial PETases with unique structural adaptations at their active sites, expanding the known structural and sequence landscape (O. Turak).[21] These marine-derived PET hydrolases offer promising potential for developing environmentally robust whole-cell catalysts when paired with similarly resilient microbial hosts. A proposed strategy to discover novel plastic-active enzymes is by inducing a restructuring of microbial communities via selective enrichment cultures.[22] These approaches, when coupled with metagenomic analysis, have enabled the detection of previously unknown prokaryotic species and enzymes derived from mangrove soil that may be involved in PET depolymerization and the catabolism of its monomers (D.J. Jiménez).[23,24]

To accelerate enzyme discovery and process optimization, integrated platforms are being developed that combine high-throughput screening, digital lab notebooks, and impedance spectroscopy (A. Hergett). These systems bridge experimental and computational workflows, allowing systematic testing of enzyme variants and streamlined bioprocess development. A versatile, tunable PET degradation platform[25] has been adapted to support alternative polyesters as substrate. The incorporation of a parallelized enzyme purification protocol allows medium throughput screening without the limitations associated with lysate-based screening methods (A. Gagsteiger). Medium- to high-throughput screening technologies— including robotics and biochemical fingerprinting—are expediting the evaluation of hundreds to thousands of PETase variants. Coupling these approaches with pretreatment strategies to improve substrate accessibility will be important for bioprocess development, as well as informing the co-design of future enzymes and polymers (S. Lim).[26]

New strategies are also being explored to valorize enzymatic degradation products. Enzymatic polycondensation methods allow the synthesis of biodegradable polyesters from biobased monomers (L. Gardossi), guided by in silico models that simulate enzyme–polymer recognition[27] and predict polymer biodegradability.[28] These last methods integrate experimental data with molecular descriptors originally developed for drug-design purposes, also with the aid of multivariate statistical analysis (A. De Piero). Integrated biocatalytic strategies that combine PET-degrading enzymes with engineered microorganisms or insects show promising results for efficient PET degradation and upcycling. One approach involves transgenic *Drosophila melanogaster* capable of secreting evolved LCC in the gut, achieving partial PET depolymerization in vivo. Another strategy uses an engineered *E. coli* strain to convert PET-derived monomers into enantiopure D- and L-alanine, demonstrating the potential of microbial systems for transforming plastic waste into valuable products (G. Molla). Enzymatic cocktails are being developed to depolymerize biodegradable plastics in mixed biowaste streams, paving the way for integrated management aimed at upcycling into biofuels, bioplastics, and other value-added products (L. Favaro, W. Myburgh, D. Rocher, D. Vezzini, L. Faggian). Genome mining of a white rot fungus revealed enzymes active on PLA and thermoplastic starch (TPS), expressed in yeast to create such cocktails (D. Vezzini). During anaerobic digestion, these cocktails also enhance hydrolysis and biomethane production from TPS-Polybutylene adipate terephthalate (PBAT) commercial bags, leveraging engineered yeast strains for improved bioplastic waste processing (L. Faggian).

AI-driven protein engineering is also advancing the development of next-generation enzymes for bio-based plastic synthesis, including the redesign of polyhydroxyalkanoate (PHA) synthases to improve catalytic efficiency and enable the synthesis of alternating PLA–PHA copolymers (F. Ballabio, T. Giorgino). Computational tools supported the design of highly thermostable PLA depolymerases, which were subsequently integrated into polymer matrices during an ultra-high-temperature extrusion process. This made possible the production of self-biodegradable PLA under home-composting conditions, demonstrating how enzyme embedding can improve polymer end-of-life.[29] This work also highlights the effectiveness of public–private partnerships (N. Panel, I. André).

Enzyme kinetics emerged as a powerful link between experimental and computational efforts. Mechanistic models of interfacial enzyme reactions offer deeper insights into enzyme–substrate interactions and inform rational enzyme selection (P. Westh). While end-point measurements and progress curves remain standard for evaluating industrial potential, kinetic analyses with well-defined physical parameters are critical for mechanistic interpretation and structure–function studies. In this context, a new class of enzymes— termed *Intermediases*—was introduced. These enzymes exhibit high catalytic activity against soluble PET fragments. When combined with PET hydrolases, Intermediases significantly enhance overall PET conversion, demonstrating up to 15-fold synergistic effects. This synergy is believed to stem from substrate competition between soluble and insoluble PET species for PETase, underscoring the importance of efficient fragment turnover in optimizing plastic-degrading enzyme systems (M. Billeskov Keller).

**Challenges, Priorities, and Future Directions**

Based on the diverse topics addressed during the workshop, a list of future challenges and research priorities emerged from the three days of presentations and discussions. This compilation reflects both the current state of the field and the strategic directions desirable to advance enzymatic plastic depolymerization at scale. While many of the examples and discussions focused on PET, the insights and principles outlined below are broadly applicable to other synthetic polymers and can inform the development of generalizable strategies for enzymatic depolymerization.

**Overcoming Crystallinity.** One of the primary barriers to cost-efficient enzymatic PET degradation is its high crystallinity, typical of beverage bottles and textiles, which significantly limits enzyme accessibility and catalytic efficiency. Developing pre-treatment methods that reduce crystallinity with lower energy input than current processes is therefore critical.[3] An additional avenue involves designing novel protein scaffolds with active sites featuring enhanced accessibility for crystalline PET, potentially working synergistically with enzymes optimized for amorphous regions. This strategy could be further strengthened by investigating how PET chain conformations—such as trans–gauche isomerization—modulate enzymatic activity, substrate binding dynamics, and overall stability.

**Multiscale Substrate–Enzyme Interactions.** A deeper understanding of enzyme–substrate interactions beyond the catalytic site is essential. Advanced multiscale modeling approaches, including coarse-grained simulations, can reveal how enzymes engage with polymeric surfaces by capturing multiple binding and unbinding events. This enables the detection of strong non-catalytic interactions at non-reactive sites, which can influence enzyme positioning and substrate accessibility. The electrostatic surface potential of enzymes plays a crucial role in polymer binding; direct modeling and engineering of these surface features could modulate adsorption and orientation on the polymer surface, thereby improving catalytic efficiency. Integration of multiscale approaches would also allow quantification of the energetic cost associated with extracting individual oligomeric chains from the bulk material—a factor that remains poorly understood. Guided by these simulations, engineering enzyme surfaces to optimize such interactions and promote productive configurations may significantly enhance catalytic performance.

**Active Site Microenvironment and pH Effects.** The local pH within an enzyme's active site may differ from that of the surrounding solution—often by as much as one pH unit, as shown experimentally.[30] Modifying the electrostatic environment of the active site offers a route to improve enzyme activity under varied pH conditions, expanding their utility across different recycling contexts.

**Reduction of NaOH consumption.** Efforts in enzymatic depolymerization increasingly focus on minimizing the need for alkaline conditions, particularly by reducing sodium hydroxide (NaOH) usage. Conventional hydrolysis methods often rely on high NaOH concentrations to maintain optimal pH and enhance product solubilization. However, this approach presents environmental and economic challenges, including the need for neutralization steps and salt waste management. While using ammonium hydroxide for pH control has recently been shown to reduce acid and base consumption by over 99%,[8] advances in understanding how external pH affects enzyme performance may enable effective enzyme engineering in synergy with process optimization. This could allow polymer degradation to proceed efficiently at near-neutral pH, substantially decreasing the need for alkaline conditions. Such a shift not only reduces the environmental impact of the process but also improves its compatibility with downstream product recovery steps.

**Enzyme Kinetics and Mechanistic Insights.** Proficient PET hydrolases typically exhibit high substrate affinity (low $K_M$) but low catalytic turnover ($k_{cat}$), particularly on insoluble PET compared to soluble fragments. This underscores the importance of substrate physical state and interfacial dynamics. Advancing the field toward industrial implementation will require: i) developing sensitive methods to track enzyme–substrate intermediates and fragment formation in real time; ii) building comprehensive models (including molecular ones) that describe the full depolymerization pathway from polymer to monomers (terephthalic acid and ethylene glycol).

**Integrated Approaches to Plastic Waste.** Given the complexity and chemical diversity of plastic waste, no single solution is likely to achieve complete degradation. A multifaceted strategy that combines enzymatic and chemical recycling pathways is essential to address the wide range of polymer types and environmental conditions found in real-world waste streams. This includes exploring synergistic enzymes that act, for instance, on soluble degradation products of PET, thereby enhancing overall depolymerization efficiency.

**Need for Standardized Experimental Frameworks Across the Field.** Many studies vary in their choice of PET substrates and data reporting formats. Without consistent metrics—particularly kinetic parameters like initial degradation rates and turnover numbers—and perhaps also consistent naming of identical protein sequences across studies, it becomes challenging to compare enzyme variants or assess progress across studies. Establishing community-wide standards for experimental design and reporting is crucial to streamline enzyme development and enhance reproducibility and scalability. A standardization effort and set of guidelines have recently been published.[5,31]

As a final note, we highlight how the small-format structure of the workshop, in line with the usual CECAM format, fostered a collaborative, friendly, and relaxed atmosphere. This welcoming environment enriched scientific exchange and collaboration—something participants consistently cited as a key success of the event. We hope this model will inspire future workshops, expanding to new and emerging topics, including plastics beyond PET. These may include other hydrolysable polymers such as polyurethanes, polylactides, and polyamides, as well as more recalcitrant materials like polyethylene. Advancing these efforts will require applying key lessons learned from PET. We also emphasize that continuous innovation in the field should be coupled with, if not driven by, a deep molecular understanding of the bioprocesses involved, synergizing with AI-based engineering and process optimization. We hope this report encourages future initiatives and the formation of shared consortia to foster an open, collaborative, and bio-based plastic recycling community.

**Acknowledgments**

CECAM-IT-SISSA-SNS and Agencia Estatal de Investigación - Spanish Ministry of Science and Innovation - PID2021-127961NB-I00 funded by MICIU/AEI /10.13039/501100011033 and by FEDER, UE are acknowledged for funding the CECAM workshop "Computations meet Experiments to Advance the Enzymatic Depolymerization of Plastics One Atom at a Time" organized in Trieste, on May 6-8, 2025.  F.B. and T.G. acknowledge PRIN 2022 from the Ministero dell'Università e Ricerca, funded by the European Union — NextGenerationEU, for project BioCat4BioPol, CUP B53D23015140006.

**Competing interests**

None declared

**Author contributions**

F.C. drafted the manuscript with the help of P. B.-S.; all authors edited and commented on the final version.